\documentclass[superscriptaddress]{revtex4}
\usepackage{graphics,epsfig}
\usepackage{graphicx}
\usepackage{dcolumn}
\usepackage{amsmath,amssymb}
\usepackage{epstopdf,hyperref,subfigure}
\newcommand{\be}{\begin{equation}}
\newcommand{\ee}{\end{equation}}
\newcommand{\beq}{\begin{equation}}
\newcommand{\eeq}{\end{equation}}
\newcommand{\bea}{\begin{eqnarray}}
\newcommand{\eea}{\end{eqnarray}}

\def\be{\begin{equation}}
\def\ee{\end{equation}}
\def\ba{\begin{eqnarray}}
\def\ea{\end{eqnarray}}

\begin{document}
	\title{Non-Local Gravity Wormholes}
	\author{Salvatore Capozziello}
	\email{capozziello@unina.it}
	\affiliation{Dipartimento di Fisica "E. Pancini", Universit\`a di Napoli ”Federico II”,
		Compl. Univ. di Monte S. Angelo, Edificio G, Via Cinthia, I-80126, Napoli, Italy,
	}
 \affiliation{Scuola Superiore Meridionale,
		Largo S. Marcellino 10, I-80138, Napoli, Italy,
	}	
 \affiliation{INFN Sez. di Napoli, Compl. Univ. di Monte S. Angelo, Edificio G, Via Cinthia, I-80126, Napoli, Italy,}
	\author{Nisha Godani}
	\email{nishagodani.dei@gmail.com}
	\affiliation{Department of Mathematics, Institute of Applied Sciences and Humanities, GLA University, Mathura, 281406, Uttar Pradesh, India.}
	
	\begin{abstract}
		We consider Non-local  Gravity  in view to obtain  stable and traversable  wormhole solutions. In particular, the class of Non-local Integral Kernel Theories of Gravity, with the inverse d'Alembert operator in
  the gravitational action,  is taken into account. 
   We  obtain  constraints for the null energy condition   and derive the field equations. Two special cases for the related Klein-Gordon equation are assumed: one where the function in the gravitational action has a linear form and another one with exponential form. In each case, we  take into account  two forms for scalar fields and derive the shape functions. Asymptotic flatness and flaring-out conditions are checked. Energy conditions and  dynamics of the solutions are examined at the throat.
   The main result is that non-local gravity contributions  allow stability and traversability of the wormhole without considering any exotic matter.
	\end{abstract}
 \date{\today}
	
	\maketitle
	


	\section{Introduction}
	
	Wormholes are the hypothetical  geometric structures that link two spacetimes or two distinct regions of the same spacetime. The first   model was proposed by Einstein and Rosen as a vacuum solution of field equations and it is called the Einstein-Rosen bridge  \cite{Rosen}. This solution, derived from the Schwarzschild one, is not traversable and present a singularity.  In 1988, a wormhole solution was  presented by Morris and Thorne  in a static and spherically symmetric form having a traversable throat at the center \cite{Morris}. Both solutions were developed  in the framework of general relativity (GR) but their existence demands the presence of exotic matter, i.e. the null energy condition must be violated  in order to achieve a stable and  traversable structure.  This fact questioned a lot the viability of this kind of solutions as realistic physical systems in the strict realm of GR. 
 
 On the other hand, alternative theories of gravity could be necessary to address the  phenomenology ranging from fundamental scales up to cosmology. In fact,    various tests are continuously performed to examine the validity of GR at various energy regimes. It is found that several  shortcomings occur  both at  ultraviolet and infrared scales \cite{Babocock,Bosma,Peebles,Padmanabhan,Spergel,Goroff} so that extensions or modifications of GR seem mandatory to address issues like quantum gravity, dark matter, or  cosmological accelerated expansion.
 
	 Various approaches have been formulated in view to alleviate GR problems towards a comprehensive description of gravitational interaction at any scale. Some generalizations are defined by modifying the Hilbert-Einstein Action or adding extra scalar fields \cite{Rep}. The most natural extensions are  $f(R)$  gravity, where a general function of Ricci scalar $R$ is used in the gravitational  Lagrangian,  and scalar-tensor gravity, where scalar fields are minimally or non-minimally coupled to gravity. Other approaches are modifications invoking different principles with respect to GR as the possible  non-necessity of the Equivalence Principle or the Lorentz Invariance violation. Some of them are the so-called teleparallel theories of gravity where dynamics  is given by torsion instead of curvature \cite{Cai}, theories    the trace of stress-energy tensor enters as a field  into the field equations, approaches where  non-metricity is supposed to lead dynamics. For  possible discussions on these approaches,  see, e.g.  Refs. \cite{Capozziello3,Felice,Nojiri2,Agostino,Bengochea,Linder,Bajardi,Frusciante,Anagnostopoulos,Capozziello5,Harko, Lavinia, Equivalence,Trinity, Anchordoqui,Papantonopoulos,Richarte,Kanti,Bronnikov1,Bronnikov2,Myrzakulov}.
  
  In most of these formulations,  it is possible to search for wormholes as  solutions for the dynamics \cite{Samanta,Yousaf,Elizalde1,Elizalde2,nisha,Godani,Nisha3,Godani88,DeBenedictis,Harko13,Rosa,Golchin,Godani2,Samanta2,Godanimpla,Godani5,Godani3,Godani4,Samanta3,nisha21,godani21,Mishra1,Nandi, Banerjee, Vittorio1,Vittorio2,Vittorio3}, but, in any case, the problems of stability and traversability of the structures have to be considered. In other words, the issue is if the problem of the violation of energy conditions  by standard fluid matter can be overcome by considering  geometric structures related to the further degrees of freedom  besides of them of GR.

  In \cite{Mimoso1,Mimoso2}, energy conditions are discussed in extended theories of gravity showing that possible violations, related to standard matter fluids, can be restored considering all the contributions to the effective stress-energy tensor acting as source in the field equations. For the consistency of wormholes, this means that viable solutions can be achieved without searching for exotic matter but considering geometric corrections \cite{Harko13}.

  After a rapid inspection of literature, it is easy to conclude that there are many possibilities to realize wormholes without considering exotic matter which, at the moment, is not experimentally probed due to no final result on the existence of  further particles over the Standard Model.  With this status of art, it seems reliable to consider the standard observed matter and search for solutions where geometric corrections can have a main role.
It is worth noticing that effective corrections implemented for defining  modified or extended theories of gravity can consist of local fields (see e.g. \cite{Rep,Nojiri1,Capozziello2}). Hence, these corrections  are realized through local actions. However, any  theory of physics should be formulated at  non-local level to be in agreement with prescriptions of quantum mechanics \cite{Barvinsky1,Barvinsky2,Buoninfante1,Buoninfante2,Mazumdar1,Mazumdar2,Mazumdar3}.
In particular,  at the one-loop level, a dynamical non-locality emerges from  the effective actions of all fundamental interactions \cite{Barvinsky3}. 

In this framework, the non-local theories of gravity have been taken into account to describe the gravitational interactions  towards a final formulation of quantum gravity \cite{Modesto1}. These theories can be divided into two main classes. Infinite derivative theories of gravity (IDGs) \cite{Modesto,Briscese} and integral kernel theories of gravity (IKGs) \cite{Deser,Maggiore,Mancarella,Nesseris,Belgacem}. See 
\cite{Acunzo,Capozziello6} for a discussion. 

The main features emerging in adopting non-local corrections in the gravitational action are the possibility to regularize and renormalize the theory at UV scales \cite{Modesto1}, the possibility to achieve  accelerated  cosmic expansion at IR scales \cite{Deser,Acunzo,Capozziello7}, and, finally, the emergence of characteristic lengths, related to the auxiliary scalar fields, which could be useful to address large-scale structure \cite{Filippo}, galactic systems \cite{Kostas}, or high-energy phenomena \cite{Gaetano}.

From a phenomenological point of view, these characteristics are hints towards a connection between quantum and classical behavior of gravitational field and one could state that non-locality, as a natural feature of gravity,  finally emerges as effective lengths in the weak-field limit.

With these considerations in mind, one can wonder if searching for wormhole solutions in the non-local gravity (NLG) context could be a straightforward way to fix the problems of stability and traversability without exotic matter.  In fact, in NLG, geometric corrections naturally emerge and can cure violations of energy conditions related to standard perfect fluids.
 
 In this work, we want to search for wormhole solutions derived in the context of NLG. In particular, we are going to consider  IKGs. These include the integral kernels of differential operators such as the inverse of the d'Alembert operator which gives rise to non-local contributions.  We will see that characteristic lengths related to NLG 
 allow stable and traversable wormhole solutions. 

 The layout of the paper is the following. Sec. II is devoted to a summary of NLG showing as gravitational field equations are improved by non-local corrections. In Sec. III, the problem of wormholes  in NLG context is considered.  Specific solutions are discussed in Sec.IV.  In particular, we discuss how physically viable forms of non-local corrections give rise to stability and traversability conditions.
 Discussion and conclusions are drawn in Sec.V.
	
	\section{Non-Local Gravity}
	
	The presence of non-local operators in gravitational action gives rise to NLG. The Hilbert-Einstein action for non-local gravity is defined by adding terms with the non-local operator in the Ricci scalar $R$. It can be given as \cite{Capozziello6} 
	\begin{eqnarray}\label{action}
	S=\int d^4x\sqrt{-g}\Big\{\frac{R}{2\kappa}[1+f(\Box^{-1}R)]+L_m\Big\},
	\end{eqnarray}
	where $g$ denotes the determinant of the metric tensor, $k=8\pi G$, $\Box\equiv \triangledown^{\nu}\triangledown_{\nu}$ is the d'Alembert operator, $f$ is a generic  function of non-local terms and $L_m$ denotes a Lagrangian density of matter fields. Following the procedure in \cite{Odi}, 
	the localized form of the action \eqref{action} can be written as 
	\begin{eqnarray}\label{action1}
	S=\int d^4x\sqrt{-g}\Big\{\frac{1}{2\kappa}[R(1+f(\chi)-\zeta)-\triangledown_{\alpha}\zeta\triangledown^{\alpha}\chi]+L_m\},
	\end{eqnarray}
	where $\chi$ and $\zeta$ are auxiliary scalar fields derived from the localization procedure and the application of the Lagrange multipliers method (see also \cite{Capozziello6}). The variation of Eq. \eqref{action1} with respect to $\zeta$ gives 
	\begin{equation}
	\Box \chi -R=0.
	\end{equation}
	This implies $\chi=\Box^{-1}R$.
	By introducing  the further scalar field $\psi\equiv f(\chi)-\zeta$, the action \eqref{action1} takes the form
	\begin{eqnarray}\label{action2}
	S=\int d^4x\sqrt{-g}\Big\{\frac{1}{2\kappa}[R(1+\psi)-f_{\chi}(\triangledown \chi)^2+\triangledown_{\alpha}\psi\triangledown^{\alpha}\chi]+L_m\Big\},
	\end{eqnarray}
	where $f_{\chi}\equiv \partial f/\partial \chi$. It is worth noticing that the NLG action \eqref{action} has been reduced to a non-minimally coupled scalar-tensor gravity model \cite{Capriolo1,Capriolo2,Capriolo3} and non-local terms results in effective scalar fields.
 
	The variation of Eq. \eqref{action2} with respect to metric tensor $g_{\mu\nu}$ gives the field equations 

\begin{equation}\label{fe}
		R_{\mu\nu}(1+\psi)-\frac{1}{2}g_{\mu\nu}[R(1+\psi)-f_{\chi}(\triangledown \chi)^2+\triangledown_{\sigma}\triangledown^{\sigma}\chi-2\Box \psi]+\triangledown_{\mu}\psi\triangledown_{\nu}\chi-\triangledown_{\mu}\triangledown_{\nu}\psi-
	f_{\chi}\triangledown_{\mu}\chi\triangledown_{\nu}\psi=\kappa T_{\mu\nu}
	\end{equation} 
	and the variation of \eqref{action2} with respect to $\chi$ gives the Klein-Gordon equation
 
	\begin{eqnarray}\label{box}
	\Box \psi-f_{\chi\chi}(\triangledown\chi)^2+2Rf_{\chi}=0.
	\end{eqnarray}
 From Eq. \eqref{fe}, the effective stress-energy tensor is given by
	\begin{equation}
	T_{\mu\nu}^{eff}=\frac{1}{1+\psi}\Bigg[T_{\mu\nu}+\frac{1}{\kappa}\Big[\frac{1}{2}\{-f_{\chi}(\triangledown \chi)^2+\triangledown_{\sigma}\psi\triangledown^{\sigma}\chi-2\Box  \psi\}g_{\mu\nu}-\triangledown_{\mu}\psi\triangledown_{\nu}\chi+
	\triangledown_{\mu}\triangledown_{\nu}\psi+f_{\chi}\triangledown_{\mu}\psi\triangledown_{\nu}\chi\Big]\Bigg].
	\end{equation}
 As we shall discuss below,
	the function $f(\chi)$ and the scalar field $\zeta$ can determine the features   of  wormhole solutions. Specifically, we shall explore  traversability and stability of wormholes for specific forms of these functions.

	\section{Wormholes in Non-Local Gravity}
	
	Now, we want to study the effects of NLG terms on  static and spherically symmetric metric representing  wormhole solutions  given by the metric
	\begin{equation}\label{metric}
	ds^2=-e^{2\Phi(r)}dt^2+\frac{dr^2}{1-\frac{b(r)}{r}}+r^2(d\theta^2+\sin^2\theta d\phi^2),
	\end{equation}
	where $r$ is the radial coordinate and $r_0$ the throat radius. The angles $\theta$ and $\phi$ are the  angular coordinates such that $\theta \in[0,\pi]$ and $\phi\in[0,2\pi]$. For  the throat radius, the conditions  $r_0>0$, $r_0\leq r \leq \infty$ have to hold. The function $\Phi(r)$ is  the {\it redshift function} and it is responsible for the gravitational redshift. 	The function $e^{2\Phi(r)}$ should be finite everywhere to avoid the horizons and then singularities. The function $b(r)$ is called  {\it shape function}. It  gives the shape of a wormhole. To be a viable solution, the shape function must fulfill the following conditions:
 
	\begin{enumerate}
	\item It has to be  $b(r_0)=r_0$ at the throat, i.e. for $r=r_0$;
	
	\item   $b'(r)<1$ at $r=r_0$, where the prime  indicates the derivative with respect to $r$;
	
	\item  $\frac{b(r)-rb^{'}(r)}{b^2(r)}>0$ for $r>r_0$ (flaring out condition);
	
	\item  $\frac{b(r)}{r}<1$  for  $r>r_0$;
	
	\item $\lim\limits_{r\rightarrow \infty}\frac{b(r)}{r}= 0$.
	\end{enumerate}
 
	In GR, the null-energy condition (NEC) at the throat is violated, being $T_{\mu\nu}k^{\mu}k^{\nu}<0$ at $r=r_0$ with $T_{\mu\nu}$ the stress energy tensor of perfect fluid matter. On the other hand, the above NLG gives the following constraint 
	\begin{equation}\label{constraint}
	T_{\mu\nu}^{eff}k^{\mu}k^{\nu}\Bigg|_{r_0}=\frac{1}{1+\psi}\Bigg[T_{\mu\nu}k^{\mu}k^{\nu}-\frac{1}{\kappa}\Big[\triangledown_{\mu}\psi\triangledown_{\nu}\chi k^{\mu}k^{\nu}-\triangledown_{\mu}\triangledown_{\nu}\psi k^{\mu}k^{\nu}-f_{\chi}\triangledown_{\mu}\psi\triangledown_{\nu}\chi k^{\mu}k^{\nu}\Big]\Bigg]\Bigg|_{r_0}
	<0.
	\end{equation}
	Assuming $1+\psi>0$, Eq. \eqref{constraint} gives
	\begin{equation}\label{constraint1}
	0<T_{\mu\nu}k^{\mu}k^{\nu}\Big|_{r_0}<\frac{1}{\kappa}\Big[\triangledown_{\mu}\psi\triangledown_{\nu}\chi k^{\mu}k^{\nu}-\triangledown_{\mu}\triangledown_{\nu}\psi k^{\mu}k^{\nu}-f_{\chi}\triangledown_{\mu}\psi\triangledown_{\nu}\chi k^{\mu}k^{\nu}\Big]\Big|_{r_0}\,,
	\end{equation}
 allowing the viability of wormhole solution also for standard matter \cite{Harko13}.
	
	From the metric \eqref{metric}, Einstein's field equations are obtained as
	\begin{eqnarray}\label{fe1}
	&&\frac{1}{2r^2}[(-4r+5b(r))\psi^{'}-b'(r)(2+2\psi+r\psi^{'})+2r(-r+b(r))\psi^{''}]+\frac{1}{2}f_{\chi}\chi^{'2}+\nonumber\\
	&&\frac{1}{2}\psi^{'}\chi^{'}\Big(1-\frac{b(r)}{r}\Big)^{-1}=\kappa \rho
	\end{eqnarray}
	
	\begin{eqnarray}\label{fe2}
	&&\frac{(1+\psi)}{r^3}(b(r)-2r^2\phi'(r)+2rb(r)\phi^{'})-\frac{1}{2}f_{\chi}\chi^{'2}+\Box \psi++\frac{\psi}{2r^2}(rb'(r)-b(r))+\nonumber\\
	&&\Big(1-\frac{b(r)}{r}\Big)(\psi^{''}-\frac{1}{2}\psi^{'}\chi^{'}-f_{\chi}^{'}\psi^{'}) =\kappa p_r
	\end{eqnarray}
	
	\begin{eqnarray}\label{fe3}
	&&\frac{(1+\psi)}{2r}[b(r)(-1+r\phi^{'}(r)+2r\phi^{'2}(r)+2r^2\phi^{''}(r))+r(b'(r)(1+r\phi^{'}(r))-\nonumber\\
	&&2r(\phi^{'}(r)+r\phi^{'2}(r)+r\phi^{''}(r)))]-\frac{1}{2}\Big(-f_{\chi}\chi^{'2}+(\psi^{'}\chi^{'}+\frac{\psi^{'}}{r})\Big(1-\frac{b(r)}{r}\Big)-2\Box \psi\Big)\times \nonumber\\ &&\Big(1-\frac{b(r)}{r}\Big)=-\kappa p_t.
	\end{eqnarray}
	These equations involve the unknown function $f(\chi)$, the scalar fields $\chi$ and $\psi$, and $\rho$, $p_r$ and $p_t$ which are the matter-energy  density, the radial and tangential pressure respectively. 
 
 We will see that the wormhole shape, as well as traversability and stability conditions,  depend on the NLG contributions entering the above dynamics.

	\section{ Wormhole Solutions}

	Solutions of field equations \eqref{fe1}-\eqref{fe3} are obtained for specific forms of function $f$ and scalar fields $\chi$ and $\zeta$. From Eqs. \eqref{fe} and \eqref{box}, the Klein-Gordon equation is given by 
	\begin{eqnarray}\label{KG}
	\chi^{'2}\Big[-\frac{f_{\chi\chi}}{2f_{\chi}}(1+\psi)+2f_{\chi}\Big]+\box \psi \Big[\frac{1+\psi}{2f_{\chi}}+3\Big]-(1+f_{\chi})\psi^{'}\chi^{'}\Big(1-\frac{b(r)}{r}\Big)=\kappa T.
	\end{eqnarray}
	We can  assume 
	\begin{eqnarray}
	1+f_{\chi}=0,
	\end{eqnarray}
	which gives $f=-\chi+c$. For simplicity, the arbitrary constant $c$ can be set to zero. So, we have $f=-\chi$. A second option is an exponential correction of the form 
	\begin{eqnarray}
	f_{\chi}=e^{\gamma \chi}\,.
	\end{eqnarray} 
  It  gives $f=\frac{e^{\gamma \chi}}{\gamma}+c_1$. A zero value for the  arbitrary constant can be assumed , i.e.  $c_1=0$, so that $f=\frac{e^{\gamma \chi}}{\gamma}$.  
 
 Some considerations are in order for these two choices. First of all, both of them can be derived requiring the existence of Noether symmetries for the action \eqref{action}. This means, as discussed in \cite{Acunzo,Capozziello6} and, in  particular, in \cite{Sebastian}, that the existence of such symmetries is a criterion to select viable NLG models. Furthermore, while the case $f=-\chi$ is the minimal non-local correction to GR, the exponential case is particularly relevant in view of {\it super-renormalizability} of NLG as discussed in detail in \cite{Modesto1,Modesto}. 
	
	Now, for these forms of function $f$, the field equations \eqref{fe1}-\eqref{fe3} can be solved. These equations involve scalar fields $\chi$ and $\zeta$, shape function $b(r)$, redshift function $\phi(r)$, energy density $\rho$ and pressures  $p_r$ and $p_t$.

 We can take $\phi(r)=0$ and $\zeta=\zeta_0(\frac{r_0}{r})^{\beta}$, where $\zeta_0$ and $\beta$ are constants. We have two choices for the scalar field $\chi$: $(i)$ $\chi=\chi_{0}(\frac{r0}{r})^{\alpha}$ and $(ii)$  $\chi=\chi_0e^{\alpha r}$, where  $\chi_0$ and $\alpha$ are constants. Thus, we can explore the wormhole solutions in two cases with two sub-cases for each case. They are the following: Case I(a): $f=-\chi, \quad \chi=\chi_{0}(\frac{r0}{r})^{\alpha}, \quad  \zeta=\zeta_0(\frac{r_0}{r})^{\beta}$; Case I(b):
	$f=-\chi, \quad \chi=\chi_0e^{\alpha r}, \quad  \zeta=\zeta_0(\frac{r_0}{r})^{\beta}$; Case II(a):	$f=\frac{e^{\gamma \chi}}{\gamma}, \quad \chi=\chi_{0}(\frac{r0}{r})^{\alpha}, \quad  \zeta=\zeta_0(\frac{r_0}{r})^{\beta}$ and Case II(b): $f=\frac{e^{\gamma \chi}}{\gamma}, \quad \chi=\chi_0e^{\alpha r}, \quad  \zeta=\zeta_0(\frac{r_0}{r})^{\beta}$.\\
	
	\noindent
	In each case, the field equations \eqref{fe1}-\eqref{fe3} are solved numerically for $b(r)$ and necessary conditions are examined. Then the energy conditions, namely the null energy condition (NEC), the weak energy condition (WEC) and the strong energy condition (SEC) are examined in each case for those values of parameters  giving  wormholes with asymptotic flatness at infinity and  satisfying the flaring out condition. These requirements allow to state if we are considering physical gravitational fields related to the wormhole.
	
Let $p$ denote the average pressure of radial and tangential matter pressures which   can be defined as 
	\begin{eqnarray}
	p(r)=\frac{1}{3}[p_r(r)+2p_t(r)]\,.
	\end{eqnarray}
The NEC is satisfied if $\rho+p\geq 0$; WEC is satisfied if $\rho>0$ and $\rho+p\geq 0$; and  SEC is satisfied if $\rho+p>0$ and $\rho+3p\geq 0$.

	Furthermore, a traversable wormhole has to be stable. This additional property  is related to  the stability of the fluid inside the wormhole. We have to define the  adiabatic sound speed  given by 
	\begin{eqnarray}
	c^2=\frac{dp}{d\rho}.
	\end{eqnarray}
	It has to be $0<\frac{dp}{d\rho}<1$ in order to have a physical perfect fluid. With this condition, the fluid inside the wormhole geometry is stable. In other words, the wormhole solution is stable. 
	
	Besides energy and stability conditions, the properties of wormholes can be defined in terms of the shape function. In the present case, the numerical solution for $b(r)$  depends on the parameters $\alpha$, $\beta$, $\chi_0$, $\zeta_0$ in Case I
	and on the parameters $\alpha$, $\beta$, $\chi_0$, $\zeta_0$, $\gamma$ in Case II. For the sake of simplicity, $\chi_0$ and $\zeta_0$ are taken as unity. The discussion of the results is the following.
 \\
 \\

	\noindent
	\textbf{Case I(a):}
	\noindent
	$f=-\chi, \quad \chi=\chi_{0}(\frac{r0}{r})^{\alpha}, \quad  \zeta=\zeta_0(\frac{r_0}{r})^{\beta}$.\\
	\noindent
	In this case,  the shape function does not satisfy the necessary properties if $\alpha<0$. For $\alpha>0$,   wormhole solutions satisfying asymptotic flatness and flaring-out conditions can be obtained for every value of $\beta$. These solutions satisfy the null and strong energy conditions at the throat  if $\beta>0$.  For $r\geq r_0$, it has to be $\beta<0$. The sound speed $\frac{dp}{d\rho}$ has a value between 0 and 1 at the throat for every value of $\beta$. Hence, favorable solutions are achieved in this sub-case. Shape function, energy conditions, flaring-out, and stability are presented in Fig. 1. It is worth noticing that the presence of NLG term $f=-\chi$ allows stable and traversable wormholes. \\

	\noindent
	\textbf{Case I(b):}
	\noindent
	$f=-\chi, \quad \chi=\chi_0e^{\alpha r}, \quad  \zeta=\zeta_0(\frac{r_0}{r})^{\beta}$.	
	\\
	\noindent
	If $\alpha>0$,  the numerical solution for $b(r)$ does not define a wormhole. However, if $\alpha<0$,  for every value of $\beta$, the necessary conditions for $b(r)$ are fulfilled. Furthermore, the ECs for $\rho$, $\rho+p$, and $\rho+3p$ have positive values at the throat only. Therefore, the ECs are satisfied only at the throat. But the condition $\frac{dp}{d\rho}<0$ shows the instability of the throat. Thus, this case provides unstable traversable wormholes.
	\\

	\noindent
	\textbf{Case II(a):}
	\noindent
	$f=\frac{e^{\gamma \chi}}{\gamma}, \quad \chi=\chi_{0}(\frac{r0}{r})^{\alpha}, \quad  \zeta=\zeta_0(\frac{r_0}{r})^{\beta}$.\\
	\noindent
	Also in this case, for $\alpha>0$, the numerical solution for $b(r)$ defines the wormhole geometry. Now, for $\gamma<0$, if $\beta<0$, then the energy density is positive for every $r$ and ECs are also valid for $r\geq r_0$ but $\frac{dp}{d\rho}<0$ at $r_0$. Thus, this subcase gives the existence of wormhole solutions  with non-exotic matter but with instability at the throat. Results are shown in Fig. 2. If $\beta>0$, energy conditions and stability conditions are satisfied at the throat only.
	Furthermore, if $\gamma>0$, then the ECs $\rho+p$ and $\rho+3p$ are positive at the throat and $0<\frac{dp}{d\rho}<1$ for every value of $\beta$. For $\gamma>0$, the ECs and $\frac{dp}{d\rho}$ are plotted in Figs. 3 and 4.
	\\

	\noindent
	\textbf{Case II(b):}
	\noindent
	$f=\frac{e^{\gamma \chi}}{\gamma}, \quad \chi=\chi_0e^{\alpha r}, \quad  \zeta=\zeta_0(\frac{r_0}{r})^{\beta}$.	\\
	\noindent
	In this case,  $\alpha$ has to  be positive for wormholes with the desired properties. Now, if $\gamma<0$,  as $\beta<0$ decreases, the region satisfying the ECs increases. If   $\beta>0$ increases, the region satisfying the energy conditions decreases. In Fig. 5, the validation of NEC is shown in a short range of $r$ for  $\beta=1$. In Fig. 6, all NEC, WEC and SEC are satisfied for $r\in(0.2,1.8)$ with $\beta=-1$. However, the throat is not stable for any value of $\beta$.  Furthermore, if $\gamma>0$, then the energy density is positive for every $r\geq r_0$ and the terms $\rho+p$ and $\rho+3p$ are positive for  $r\leq 1.5$. This shows that the ECs are valid at and near the throat only. As for $\gamma<0$, the throat is also  not stable for $\gamma>0$.

	\begin{figure}
		\centering
		\subfigure[]{\includegraphics[scale=.6]{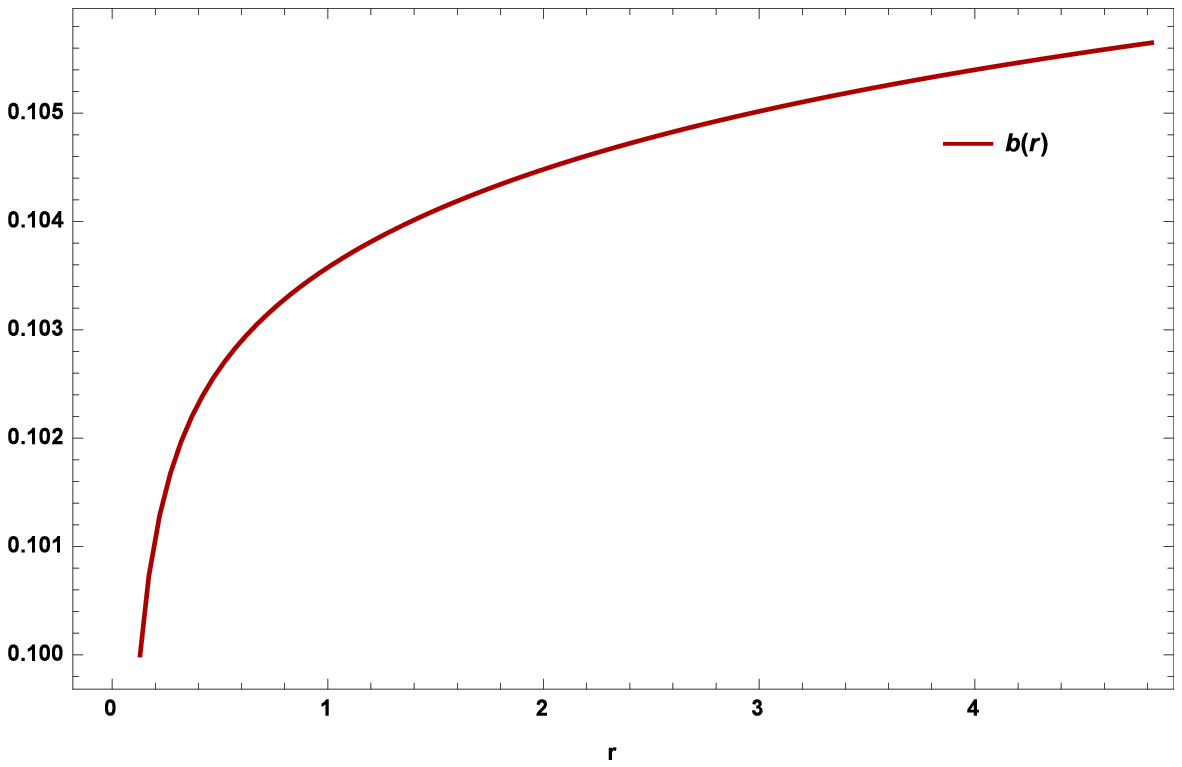}}\hspace{.5cm}
		\subfigure[]{\includegraphics[scale=.6]{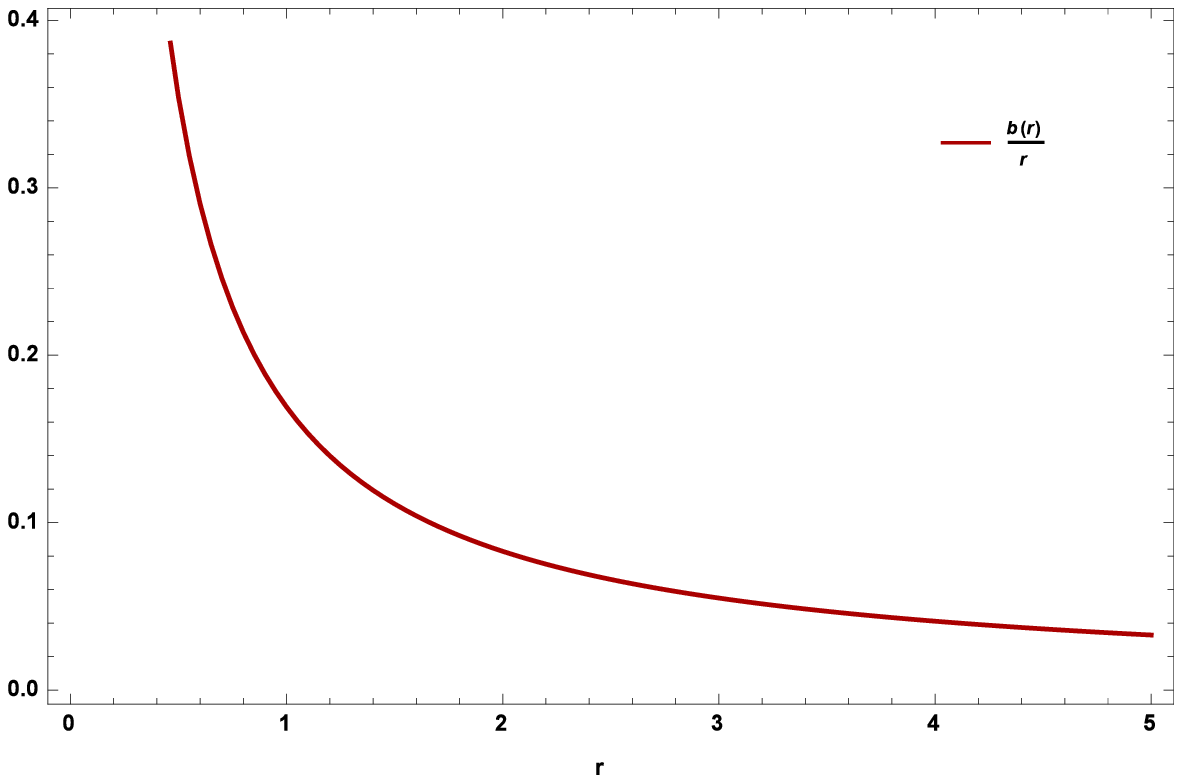}}\hspace{.5cm}\\
		\subfigure[]{\includegraphics[scale=.6]{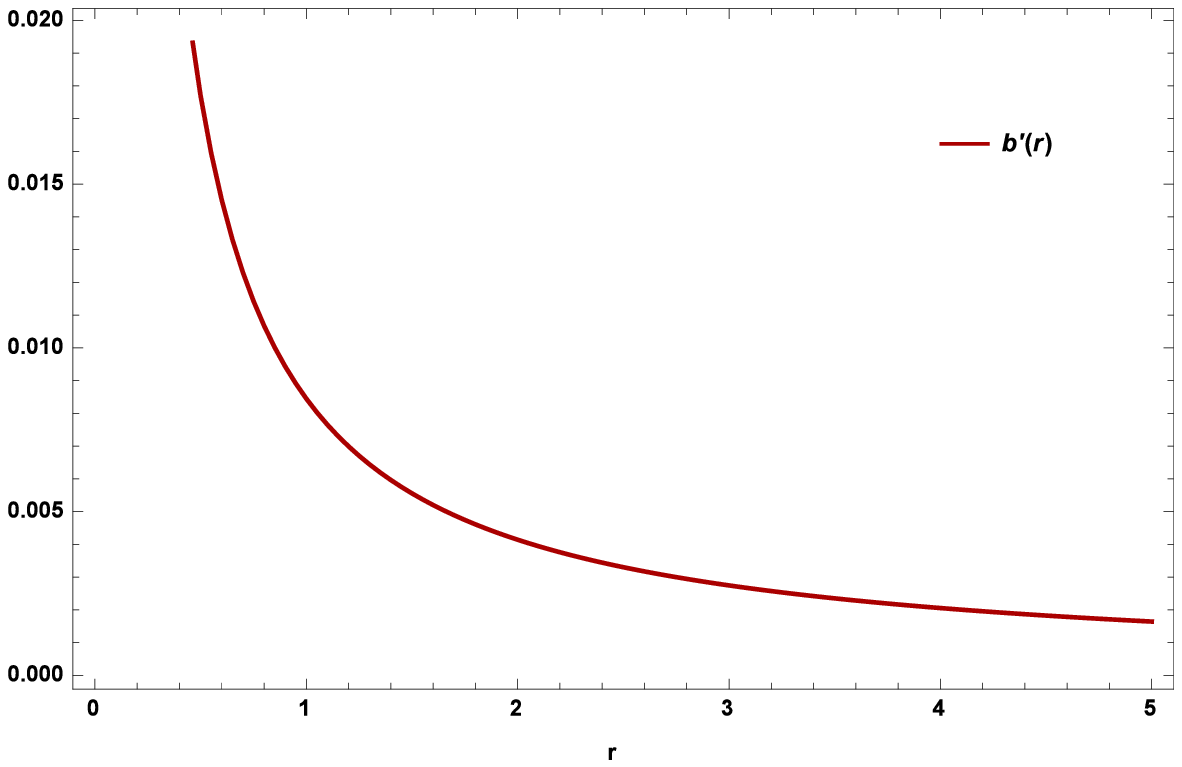}}\hspace{.5cm}
		\subfigure[]{\includegraphics[scale=.6]{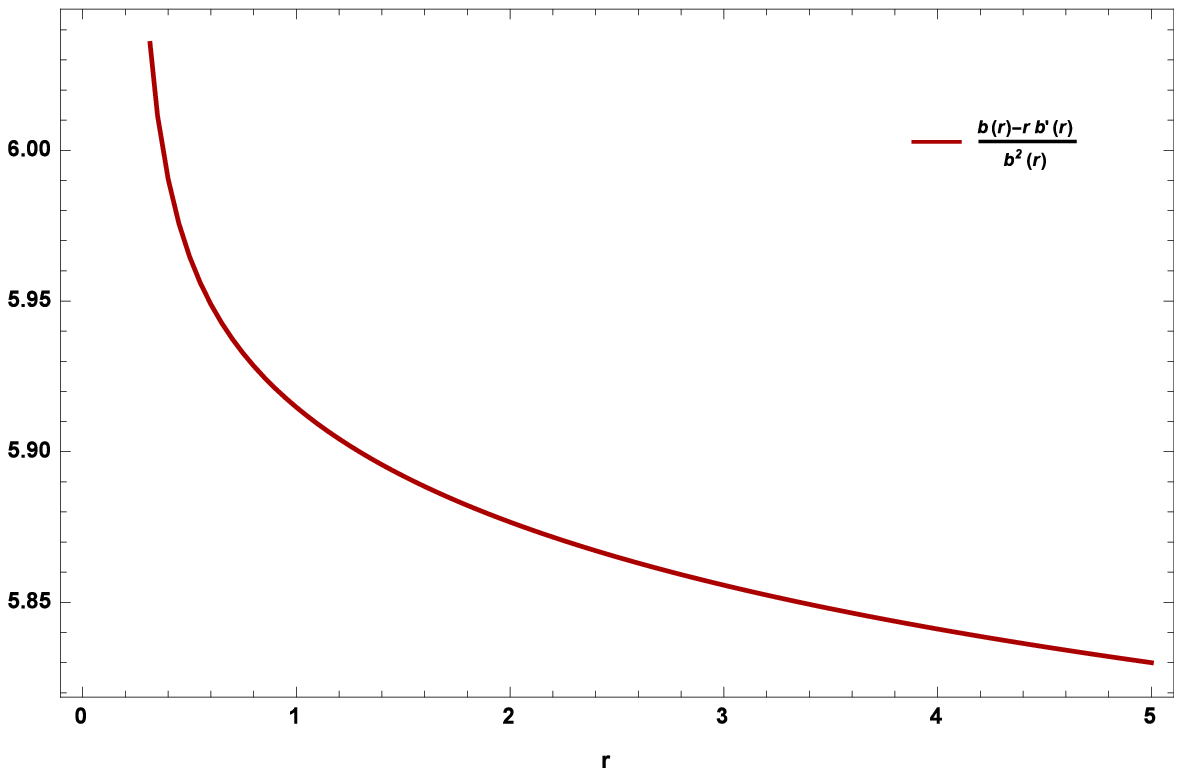}}\hspace{.5cm}\\
		\subfigure[]{\includegraphics[scale=.83]{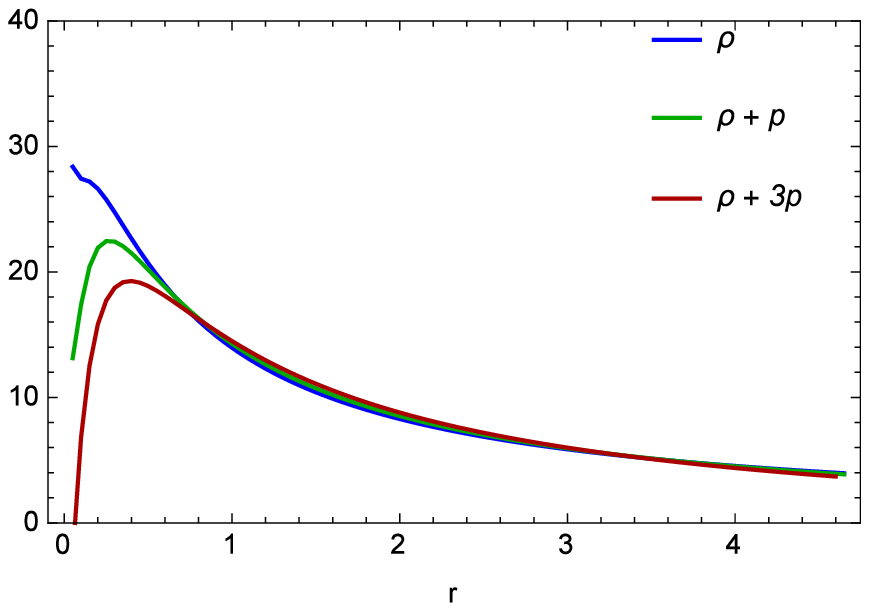}}\hspace{.5cm}
		\subfigure[]{\includegraphics[scale=.83]{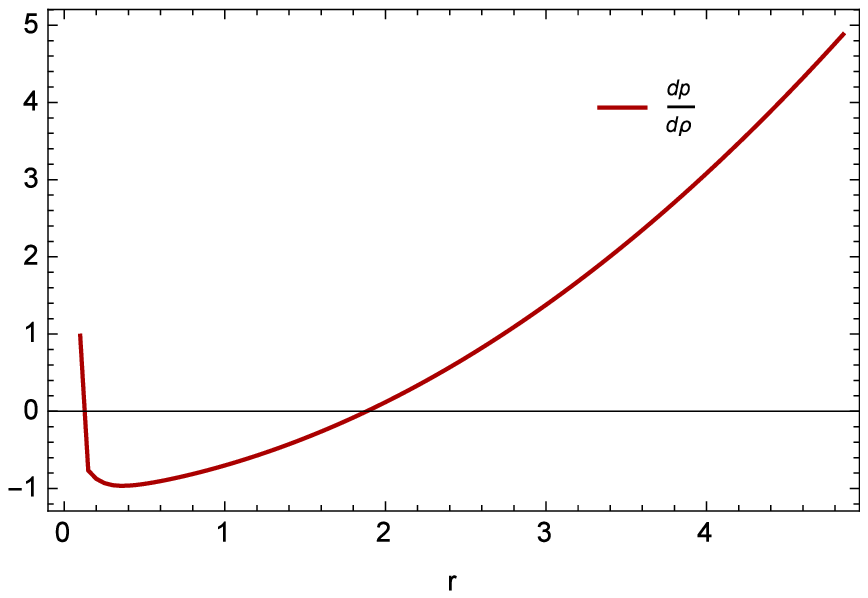}}\hspace{.5cm}
		\caption{Case I(a): The plots represent the shape function $b(r)$, $\frac{b(r)}{r}$, $b^{'}(r)$, $\frac{b(r)-rb^{'}(r)}{b^{2}(r)}$; EC terms $\rho$, $\rho+p$, $\rho+3p$ and sound speed $\frac{dp}{d\rho}$.  The values of the parameters are $\alpha=3$ and $\beta=-1$.  }
	\end{figure}
	
	\begin{figure}
		\centering
		\subfigure[]{\includegraphics[scale=.6]{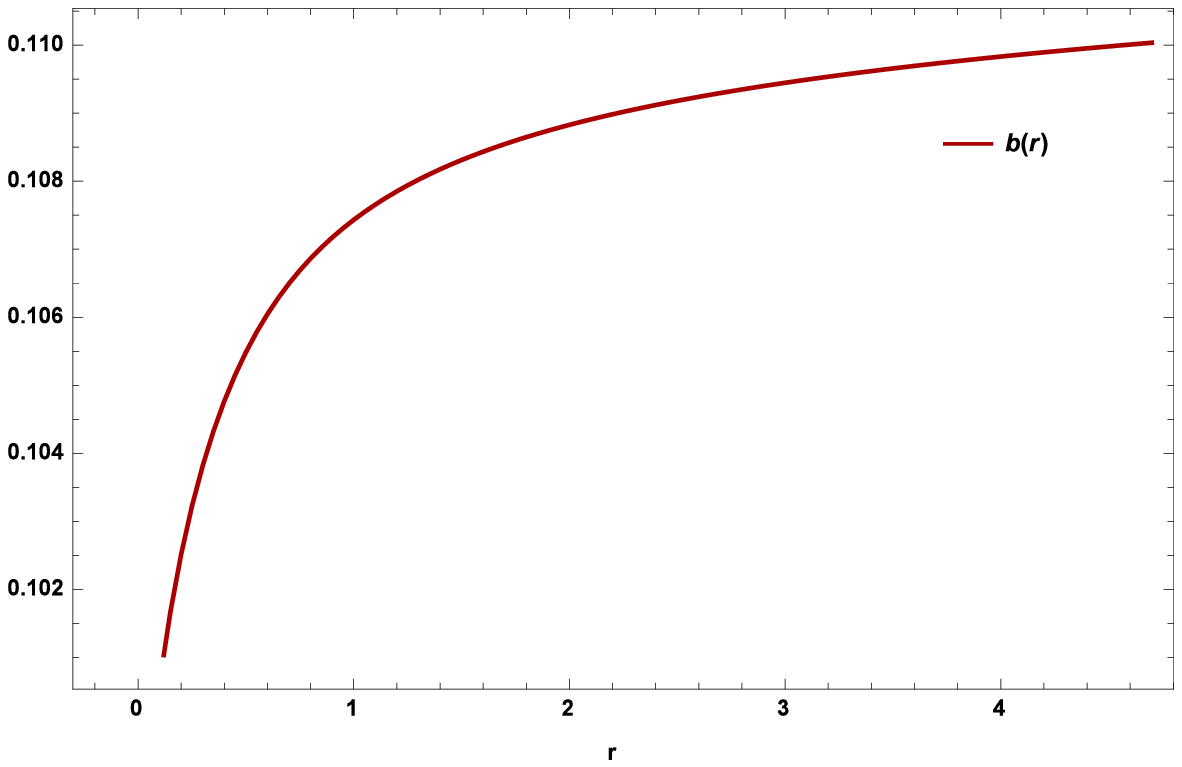}}\hspace{.5cm}
		\subfigure[]{\includegraphics[scale=.6]{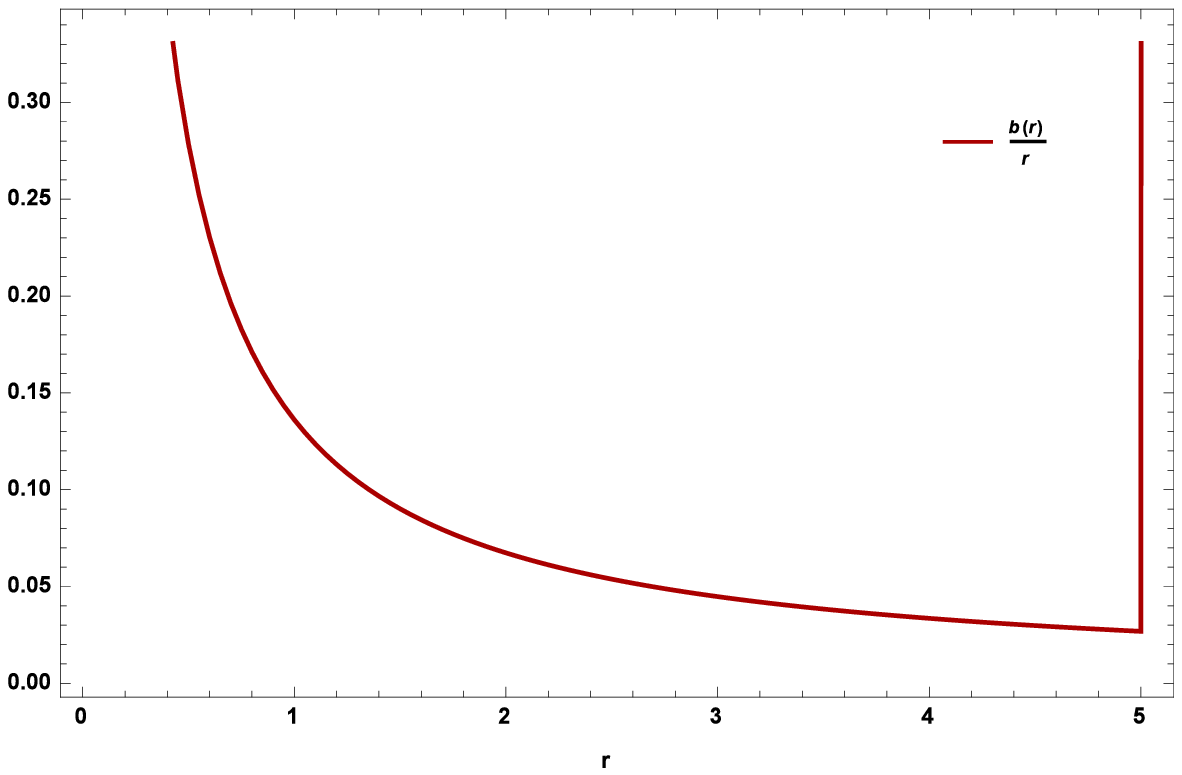}}\hspace{.5cm}\\
		\subfigure[]{\includegraphics[scale=.6]{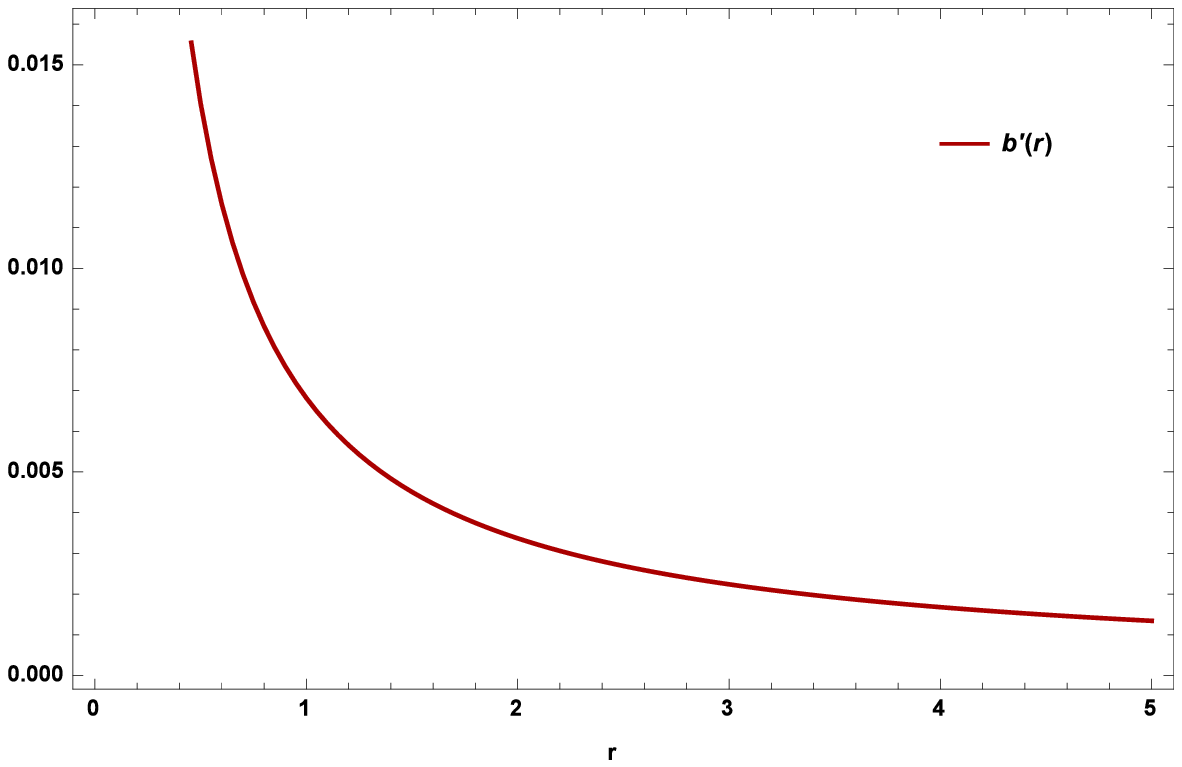}}\hspace{.5cm}
		\subfigure[]{\includegraphics[scale=.6]{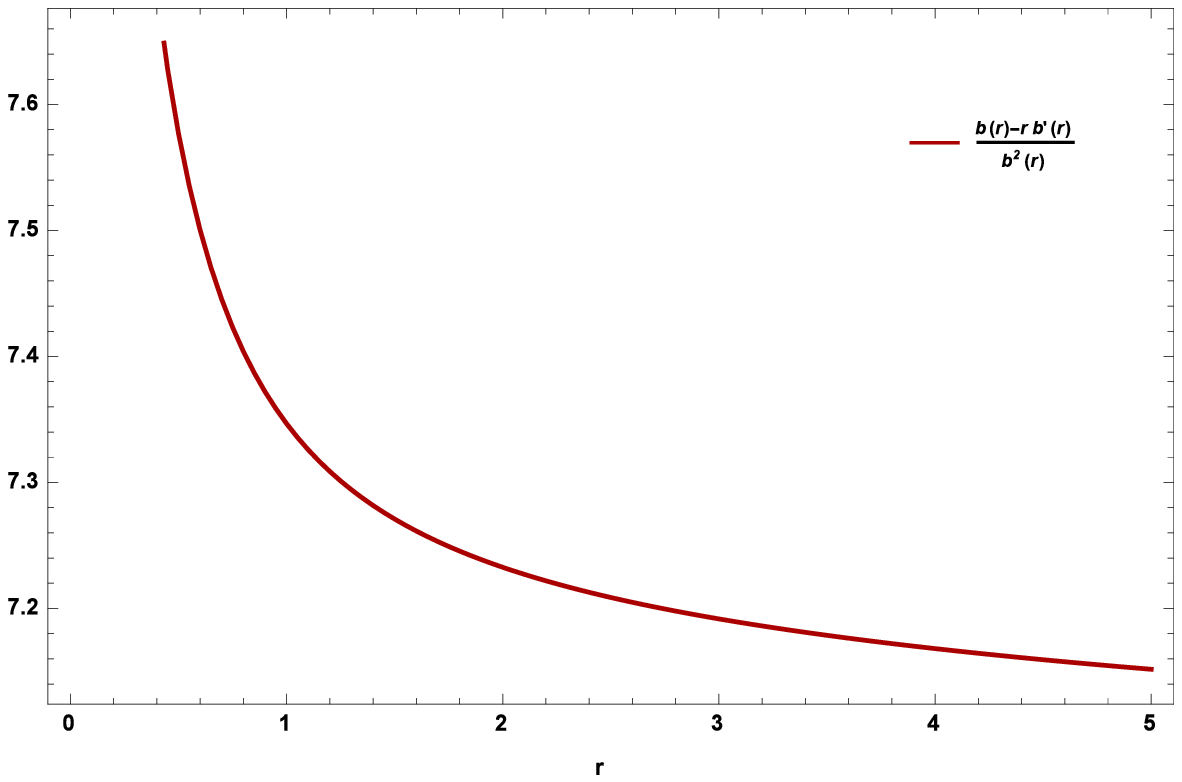}}\hspace{.5cm}\\
		\subfigure[]{\includegraphics[scale=.83]{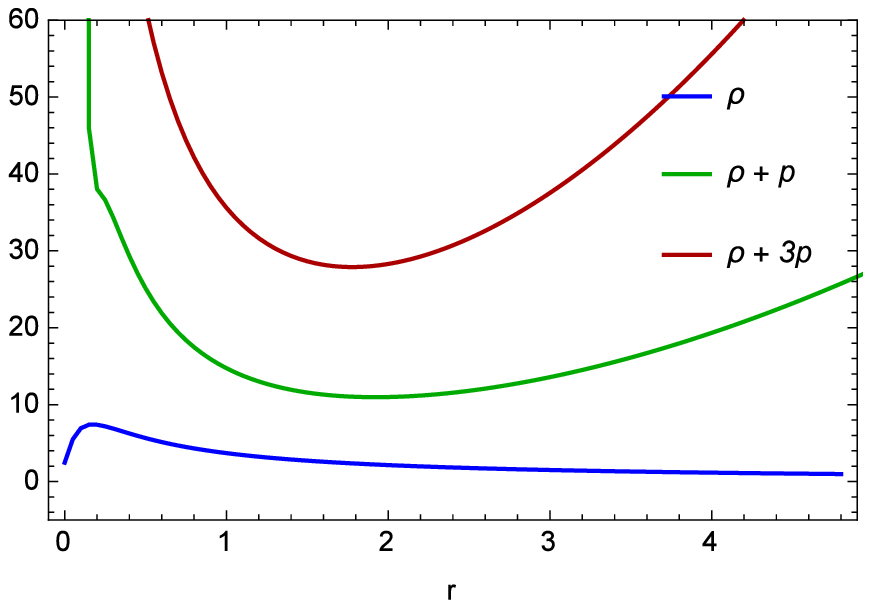}}\hspace{.5cm}
		\subfigure[]{\includegraphics[scale=.83]{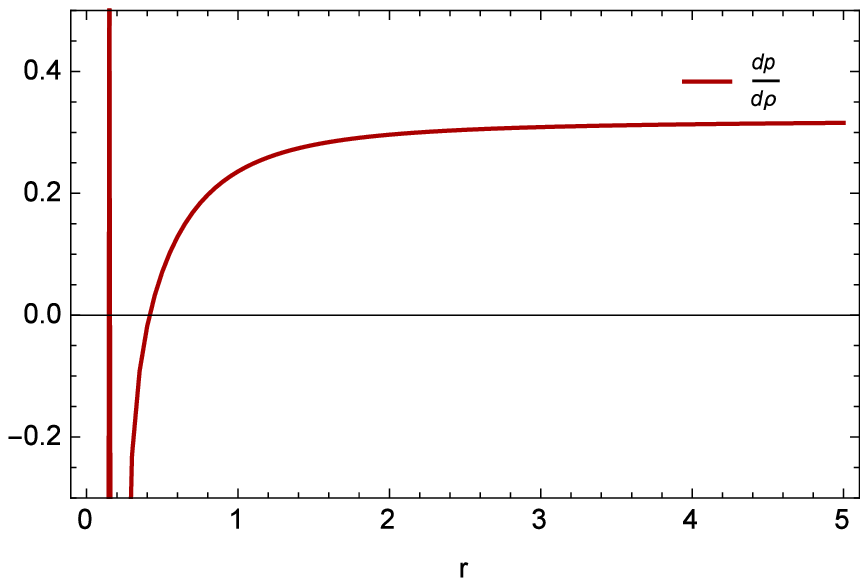}}\hspace{.5cm}
		\caption{Case II(a): In this  figure are plotted the shape function $b(r)$, $\frac{b(r)}{r}$, $b^{'}(r)$, $\frac{b(r)-rb^{'}(r)}{b^{2}(r)}$; EC terms $\rho$, $\rho+p$, $\rho+3p$ and sound speed $\frac{dp}{d\rho}$ are plotted with $\alpha=1$, $\beta=-1$ and $\gamma=-1$.}
	\end{figure}
	
	\begin{figure}
		\centering
		\subfigure[]{\includegraphics[scale=.78]{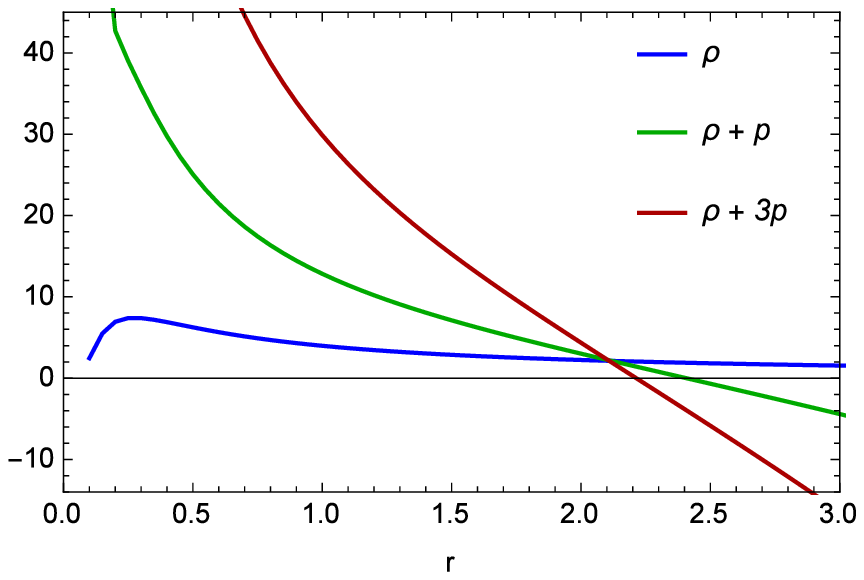}}\hspace{.5cm}
		\subfigure[]{\includegraphics[scale=.82]{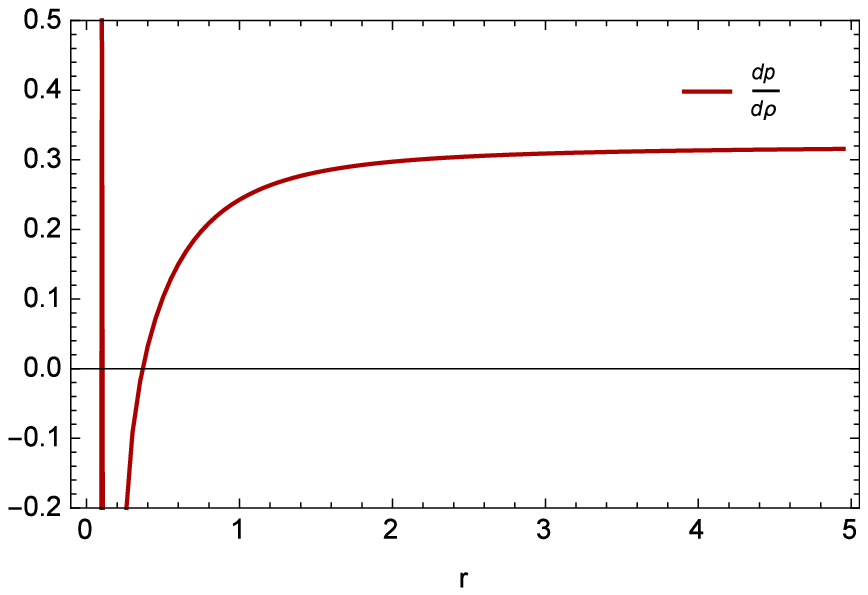}}\hspace{.5cm}
		\caption{Case II(a): In this  figure, EC terms $\rho$, $\rho+p$, $\rho+3p$ and sound speed $\frac{dp}{d\rho}$ are plotted with $\alpha=1$, $\beta=-1$ and $\gamma=1$.}
	\end{figure}
	\begin{figure}
		\centering
		\subfigure[]{\includegraphics[scale=.83]{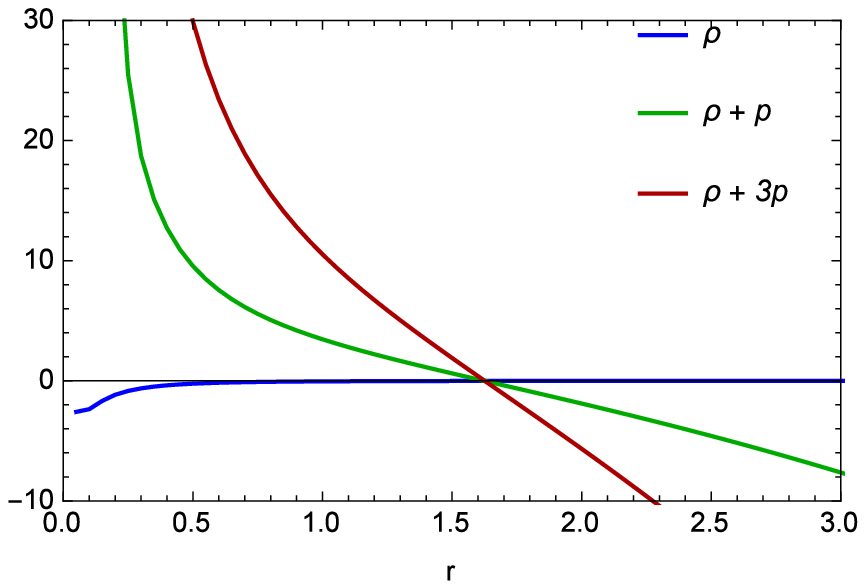}}\hspace{.5cm}
		\subfigure[]{\includegraphics[scale=.83]{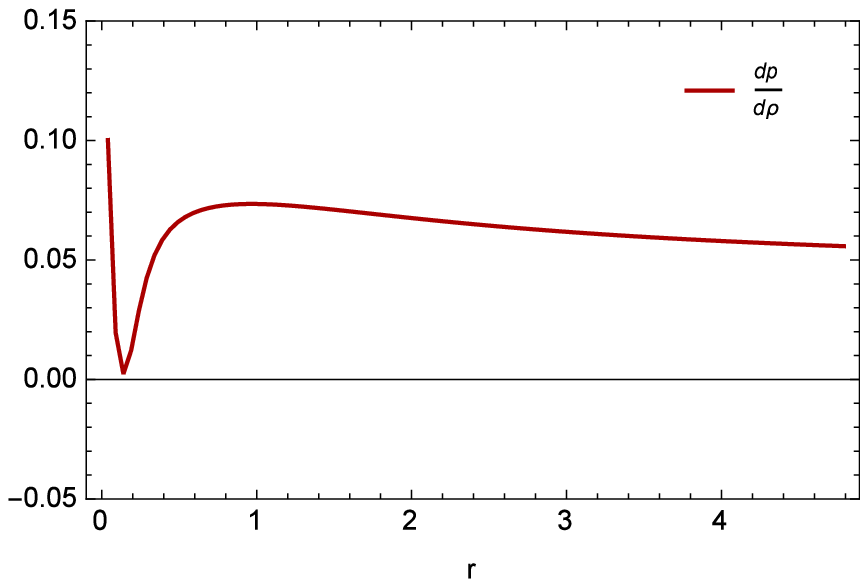}}\hspace{.5cm}
		\caption{Case II(a): In this  figure, EC terms $\rho$, $\rho+p$, $\rho+3p$ and sound speed $\frac{dp}{d\rho}$ are plotted with $\alpha=1$, $\beta=1$ and $\gamma=1$.}
	\end{figure}
	\begin{figure}
		\centering
		\subfigure[]{\includegraphics[scale=.84]{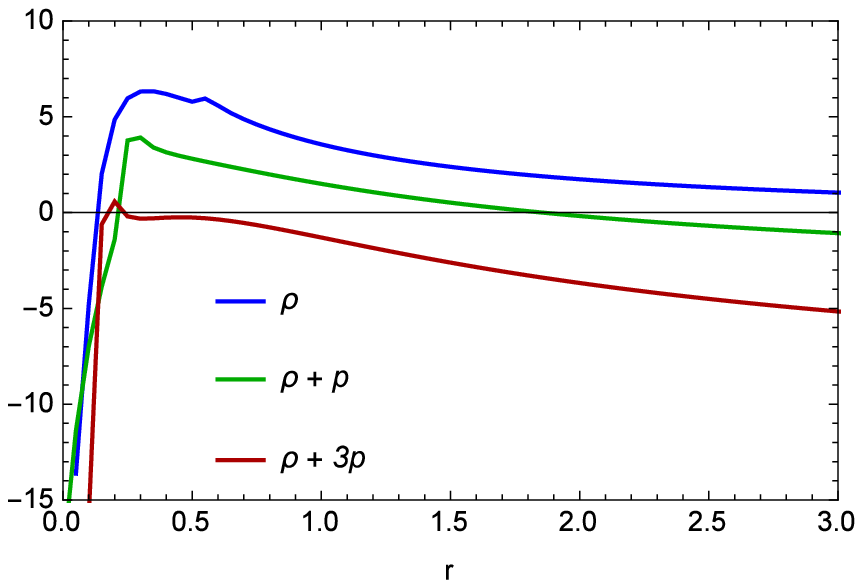}}\hspace{.5cm}
		\subfigure[]{\includegraphics[scale=.84]{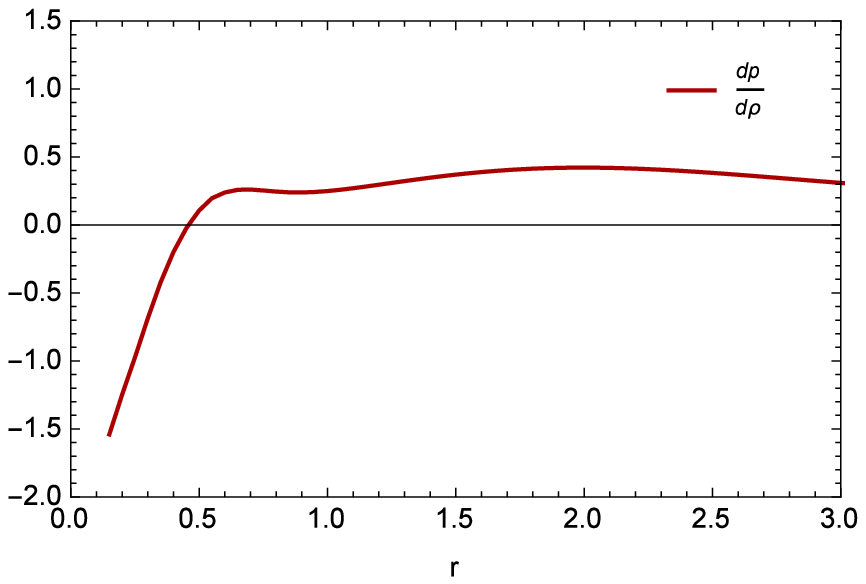}}\hspace{.5cm}
		\caption{Case II(b): Here, EC terms $\rho$, $\rho+p$, $\rho+3p$ and sound speed $\frac{dp}{d\rho}$ are plotted with $\alpha=1$, $\beta=1$ and $\gamma=-1$.}
	\end{figure}
	\begin{figure}
		\centering
		\subfigure[]{\includegraphics[scale=.84]{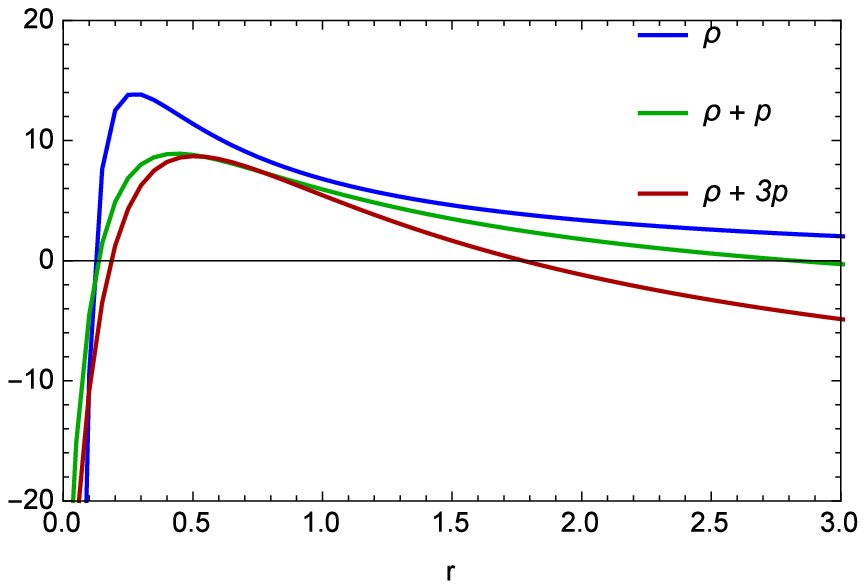}}\hspace{.5cm}
		\subfigure[]{\includegraphics[scale=.84]{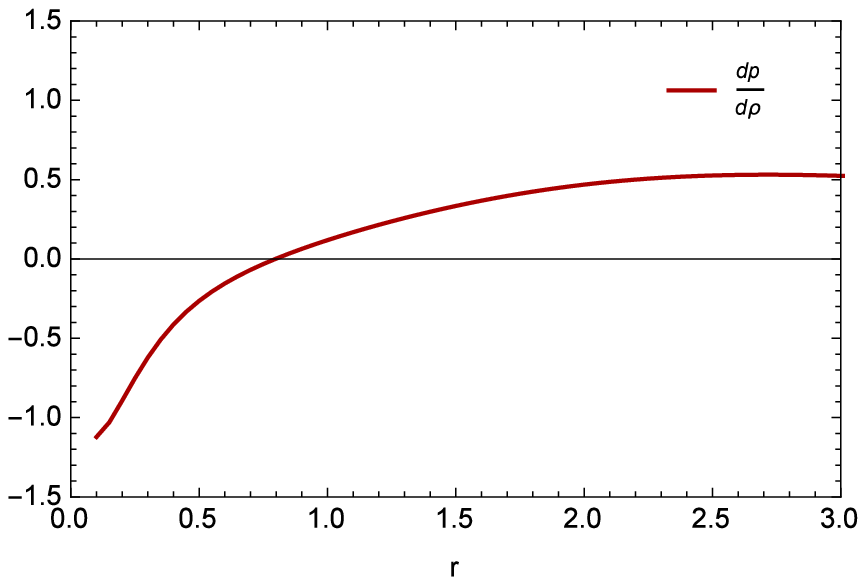}}\hspace{.5cm}
		\caption{Case II(b): In this  figure, EC terms $\rho$, $\rho+p$, $\rho+3p$ and sound speed $\frac{dp}{d\rho}$ are plotted with $\alpha=1$, $\beta=-1$ and $\gamma=-1$.}
	\end{figure}
	%
\newpage	
\section{Discussion and Conclusions}
 Wormhole solutions require exotic matter to be stable and traversable structures in the framework of GR. Due to this fact, several people excluded them as reliable astrophysical objects because, at the moment, there is no final evidence of particles, other than the Standard Model, which could constitute viable candidates for some form of matter capable of violating the energy conditions. 
 
 An important remark is necessary at this point.  As discussed in \cite{Em1}, 
  traversable wormholes can be obtained in the framework of the Einstein-Dirac-Maxwell theory where Dirac spinor and electromagnetic fields are considered. Essentially the authors achieve singularity-free configurations which are asymptotically flat. Such solutions can be naturally connected to Reissner-Nordstr\"om black holes where the mass $M$ and the electric charge $Q$ of the self-gravitating system play a major role without using phantom matter.  Apparently this kind of configurations could not exist in nature because extended charged astrophysical systems have never been detected by observations.  Furthermore, as reported in \cite{Em2},   wormhole solutions,  asymmetric with respect to  the throat and endowed with smooth gravitational and matter spinor fields, can  also be achieved in a realistic scenario.   However,  a physically relevant condition  is necessary to lead to no gravitational force experienced by a stationary observer at the throat. In \cite{Em3},  this condition is deemed  unnecessary. In conclusion,   the discussion on the realizability of wormhole solutions in the framework of GR is today extremely rich and interesting.

 In this debate, extended and modified theories of gravity can assume a major role allowing to alleviate several shortcomings of GR at UV and IR scales.  The recipe is that traversability and stability conditions for wormholes could be restored thanks to the different number of degrees of freedom  showed by several classes of these alternative to GR.

 An important case among these possibilities is  represented by NLG. Models related with this approach can be considered as a natural extension of GR towards quantum gravity because allow regularization and renormalization of gravitational effective action \cite{Modesto1,Modesto}. One of the characteristic features of NLG is that non-local terms result as effective scalar fields after localization procedures. From a phenomenological point of view, this means that non-local corrections give rise to effective lengths and masses which could cure some GR issues. 

 In this context, the viability of wormhole solutions can be considered by searching for stable and traversable structures where non-local terms play a main role.

 In this paper, we have taken into account spherically symmetric Morris-Thorne wormhole solutions satisfying NLG field equations. In particular, we considered   linear and  exponential NLG corrections motivated by the fact that such terms, in the gravitational action, are selected by the existence of Noether symmetries \cite{Acunzo, Capozziello6, Sebastian}. The result is that, in both cases, it is possible to obtain viable wormhole solutions whose stability and traversability is directly related to the presence of non-local terms. 

 In our opinion, the interest for this  result is twofold. From one hand, if observed, wormholes could have a natural explanation in the context of gravitational theories without exotic matter. On the other hand, they could be a test bed for effective theories of gravity towards quantum gravity.

 In a forthcoming paper, following the approaches presented in \cite{Vittorio1,Vittorio2,Vittorio3}, we will discuss possible observational constraints for NLG wormholes.
 
\section*{Acknowledgements}
SC acknowledges the Istituto Nazionale di Fisica Nucleare (INFN), Sezione di Napoli, \textit{iniziative specifiche}  MOONLIGHT2 and  QGSKY  for the support.

\end{document}